\documentclass[
reprint,
superscriptaddress,
amsmath,
amssymb,
showkeys,
aps,
prl,
nolongbibliography,
]{revtex4-2}

\usepackage{threeparttable}
\usepackage{setspace}
\usepackage{makecell}
\usepackage{amsmath}
\usepackage{graphicx}
\usepackage{dcolumn}
\usepackage{bm}
\usepackage{hyperref}
\usepackage[mathlines, pagewise]{lineno}
\usepackage{textcomp}
\usepackage{gensymb}
\usepackage{subfigure}
\usepackage{url}
\usepackage{xcolor}
\usepackage{soul}
\usepackage{hhline}
\usepackage[version=3]{mhchem}
\usepackage{multirow}
\usepackage[american]{babel}
\usepackage{textgreek}

\begin{document}

\title{
Crystallization Instead of Amorphization in Collision Cascades in Gallium Oxide
}

\author{Junlei Zhao} 
% \email{zhaojl@sustech.edu.cn}
\affiliation{Department of Electrical and Electronic Engineering, Southern University of Science and Technology, Shenzhen 518055, China}

\author{Javier Garc{\'i}a Fern{\'a}ndez}
%\email{j.g.fernandez@smn.uio.no}
\affiliation{Department of Physics and Centre for Materials Science and Nanotechnology, University of Oslo, PO Box 1048 Blindern, N-0316 Oslo, Norway}

\author{Alexander Azarov}
%\email{alexander.azarov@smn.uio.no}
\affiliation{Department of Physics and Centre for Materials Science and Nanotechnology, University of Oslo, PO Box 1048 Blindern, N-0316 Oslo, Norway}

\author{Ru He}
% \email{ru.he@helsinki.fi}
\affiliation{Department of Physics and Helsinki Institute of Physics, University of Helsinki, P.O. Box 43, FI-00014, Finland}

\author{{\O}ystein Prytz}
%\email{oystein.prytz@fys.uio.no}
\affiliation{Department of Physics and Centre for Materials Science and Nanotechnology, University of Oslo, PO Box 1048 Blindern, N-0316 Oslo, Norway}

\author{Kai Nordlund} 
% \email{kai.nordlund@helsinki.fi}
\affiliation{Department of Physics and Helsinki Institute of Physics, University of Helsinki, P.O. Box 43, FI-00014, Finland}

\author{Mengyuan Hua} 
% \email{huamy@sustech.edu.cn}
\affiliation{Department of Electrical and Electronic Engineering, Southern University of Science and Technology, Shenzhen 518055, China}

\author{Flyura Djurabekova} 
%\email{flyura.djurabekova@helsinki.fi}
\affiliation{Department of Physics and Helsinki Institute of Physics, University of Helsinki, P.O. Box 43, FI-00014, Finland}

\author{Andrej Kuznetsov}
% \email{andrej.kuznetsov@fys.uio.no}
\affiliation{Department of Physics and Centre for Materials Science and Nanotechnology, University of Oslo, PO Box 1048 Blindern, N-0316 Oslo, Norway}

\begin{abstract} 

Disordering of solids typically leads to amorphization, but polymorph transitions, facilitated by favorable atomic rearrangements, may temporarily help to maintain long-range periodicity in the solid state. 
In far-from-equilibrium situations, such as atomic collision cascades, these rearrangements may not necessarily follow a thermodynamically gainful path, but may be kinetically limited. 
In this Letter, we focused on such crystallization instead of amorphization in collision cascades in gallium oxide (\ce{Ga2O3}). 
We determined the disorder threshold for irreversible $\beta$-to-$\gamma$ polymorph transition and explained why it results in elevating energy to that of the $\gamma$-polymorph, which exhibits the highest polymorph energy in the system below the amorphous state. 
Specifically, we demonstrate that upon reaching the disorder transition threshold, the \ce{Ga}-sublattice kinetically favors transitioning to the $\gamma$-like configuration, requiring significantly less migration for \ce{Ga} atoms to reach the lattice sites during post-cascade processes. 
As such, our data provide a consistent explanation of this remarkable phenomenon and can serve as a toolbox for predictive multi-polymorph fabrication.

\end{abstract}

\maketitle

Notably, while amorphization in cascades is overwhelmingly common in many materials, there have been a few reports in literature confirming radiation-induced polymorphs transitions~\cite{lumpkin2008experimental, anber2020structural}.
Nevertheless, until very recently, interest in such crystallization instead of amorphization phenomenon was primarily maintained within the specialized radiation effect research community, despite the great options for tuning functional properties, however in reality limited by the challenges to demonstrate well-defined structures commonly required in technology.
This was the status quo until Azarov \textit{et al.}~\cite{azarov2022disorder} demonstrated a regularly-shaped new polymorph thin film on the top of the initial polymorph substrate, as a result of the radiation-induced disordering in gallium oxide (\ce{Ga2O3}). 
This work attracted significant attention of a broader research community, inspired by this novel opportunity to design new functionalities out of \ce{Ga2O3} polymorph stacks, potentially useful in a range of \ce{Ga2O3} technologies, from power electronics~\cite{green2022gallium, tadjer2022toward} to solar-blind ultraviolet optoelectronics~\cite{pratiyush2019advances, kim2020highly}.
In the majority of the \ce{Ga2O3} polymorph transition studies thus far, monoclinic $\beta$-\ce{Ga2O3} (12, $C2/m$) was used as the initial material, being selected as the thermodynamically stable form of \ce{Ga2O3}.
Additionally, at least four metastable \ce{Ga2O3} polymorphs (sorted as $\beta < \kappa < \alpha < \delta < \gamma$ with respect to the ascending order of the zero-strain potential energies) have been reported in literature~\cite{roy1952polymorphism, yoshioka2007structures, swallow2020influence, mu2022phase, ratcliff2022tackling, kato2023demonstration}. 

Importantly, initial reports~\cite{anber2020structural, azarov2022disorder} identified the disorder-induced polymorph as the orthorhombic $\kappa$-\ce{Ga2O3} (33, $Pna2_{1}$) accounting for its lowest -- after $\beta$-\ce{Ga2O3} -- energy also supported by the electron microscopy identification, however at that time performed along one zone-axis only; while the later works~\cite{garcia2022formation, yoo2022atomic, SFazarov2023universal, huang2023atomica, huang2023atomicb} unambiguously identified the newly formed polymorph as cubic defective spinel $\gamma$-\ce{Ga2O3} (227, $Fd\overline{3}m$) based on multiple zone-axis microscopy investigations.
Thus, even though there is a consensus for the disorder-induced $\beta$-to-$\gamma$ phase transition in \ce{Ga2O3} occurring instead of amorphization in literature, no clear explanation exists as to why the transition lifts the system energy to that of the highest among all \ce{Ga2O3} polymorphs; even though this result indicates that the process is likely to be kinetically limited.
Moreover, despite the considerable data already collected on radiation-induced polymorphism in \ce{Ga2O3}, the disorder thresholds enabling the transition have never been determined with enough precision to enable predictive modeling for potential applications. 

In this Letter, we show that the $\beta$-to-$\gamma$ phase transition occurs remarkably swiftly, governed by a sharp disorder threshold. 
Specifically, by leveraging our recently developed machine-learning (ML) molecular dynamics (MD) model~\cite{SFzhao2023complex}, we precisely pinpoint this narrow threshold range of accumulated disorder that triggers the irreversible $\beta$-to-$\gamma$ phase transition during post-cascade evolution, in excellent agreement with the experimental data. 
As such, our systematic data provide a consistent explanation for the phenomenon of crystallization instead of amorphization in \ce{Ga2O3} and can serve as a toolbox for predictive fabrication of multi-polymorphic structures.

\begin{figure*}[ht!]
 \includegraphics[width=13cm]{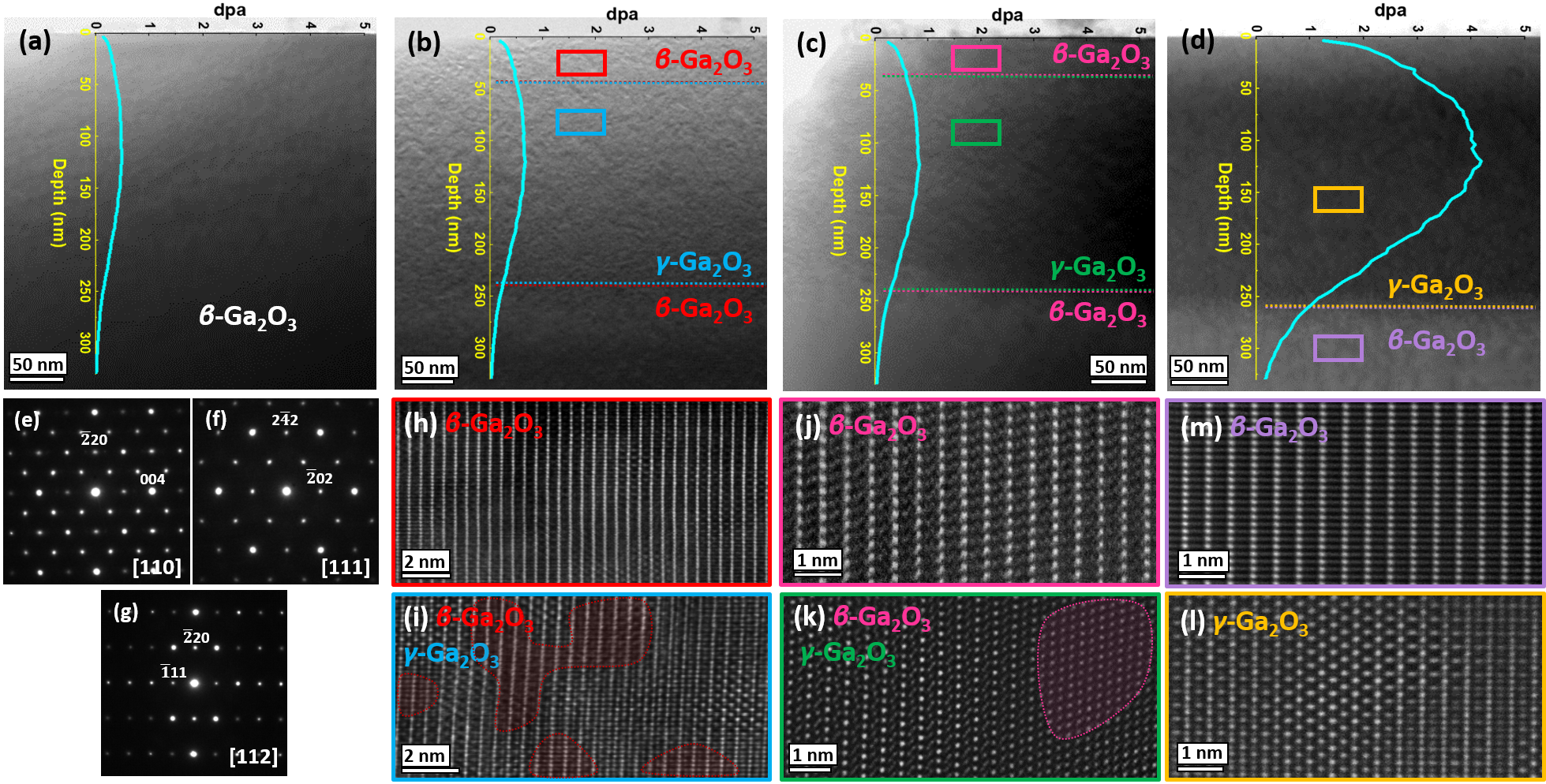}
 \caption{
    Swift $\beta$-to-$\gamma$ \ce{Ga2O3} phase transition controlled by a disorder threshold.
    Bright field STEM images showing fully implanted regions of the $\beta$-\ce{Ga2O3} samples irradiated with (a) $1.2\times10^{14}$, (b) $1.6\times10^{14}$, (c) $2.0\times10^{14}$, and (d) $1.0\times10^{15}$ $^{58}\mathrm{Ni}^{+}$ $\mathrm{cm}^{-2}$.
    The cyan curves show the corresponding damage levels in dpa unit, derived from SRIM simulations.
    The dashed lines indicate the interfaces between the $\beta$ and $\gamma$ phases.
    The colored boxes highlight the exact positions of the corresponding atomic-resolution STEM images shown in (h-l). 
    (e-g) The SAED patterns of the $\gamma$-phase collected along the $[110]$, $[111]$, $[112]$ zone axes, respectively.
    (h-l) Atomic-resolution HAADF-STEM images obtained from the samples in (b-d), linked by the colors of the boxes. 
    The images (h), (j), and (m) focus on the $\beta$-\ce{Ga2O3} regions, while the images (i), (k), and (l) focus on the $\gamma$-\ce{Ga2O3} regions.
 }
 \label{fig:exp}
\end{figure*}

Fig.~\ref{fig:exp} summarizes scanning transmission electron microscopy (STEM) data used to monitor the onset of the disorder-induced $\beta$-to-$\gamma$ phase transition and to determine the disorder threshold required for this transition to occur.
Specifically, Fig.~\ref{fig:exp}(a-d) shows bright-field STEM images of the samples implanted with $1.2\times10^{14}$, $1.6\times10^{14}$, $2\times10^{14}$ and $1\times10^{15}$ $^{58}\mathrm{Ni}^{+}$ cm$^{-2}$, corresponding to 0.50, 0.65, 0.85, and 4.15 displacements per atom (dpa) at the depth of the maximum disorder, respectively.
These are consistently with the dpa profiles (cyan curves) inserted into these panels. 
See Supplemental Material (SM) Appendix I for the experimental details and the dpa extraction from SRIM simulations.
Notably, the dashed lines in Fig.~\ref{fig:exp}(b-d) mark the positions of the established top $\beta$/$\gamma$ and bottom $\gamma$/$\beta$ interfaces, while Fig.~\ref{fig:exp}(e-g) shows characteristic selected-area electron diffraction (SAED) patterns of the $\gamma$-phase collected along the $[110]$, $[111]$, $[112]$ zone axes, respectively. 
Concurrently, Fig.~\ref{fig:exp}(h-l) shows atomic resolution HAADF-STEM images recorded from different samples and from different parts of the samples in accordance with the frame color codes, with the spots representing \ce{Ga} atoms, where the brighter ones indicates higher atomic density.
Specifically, Fig.~\ref{fig:exp}(h) and (j) correspond to the $\beta$-phase remaining above $\beta$/$\gamma$ interfaces in Fig.~\ref{fig:exp}(b) and (c), respectively, while Fig.~\ref{fig:exp}(m) is a fingerprint of the $\beta$-phase beneath the bottom $\gamma$/$\beta$ interfaces.
Notably, the visually different orientations and scales of the \ce{Ga} planes in Fig.~\ref{fig:exp}(h), (j), and (m) attribute sorely to variations in observation angle and magnification, otherwise verifying the same atomistic structure of the $\beta$-phase.
Atomic resolution images in Fig.~\ref{fig:exp}(i), (k), and (l), in turn, are taken from inside the disordered layer of the samples in Fig.~\ref{fig:exp}(b-d), respectively, clearly resolving characteristic $\gamma$-type \ce{Ga} planes.

To this end, Fig.~\ref{fig:exp} provides clear evidence that the $\beta$-to-$\gamma$ phase transition is a function of disorder, consistent with the dpa depth profiles.
Indeed, a careful inspection of the sample in Fig.~\ref{fig:exp}(a), exposed to 0.50 dpa in maximum, reveals only the $\beta$-phase characteristic features throughout the irradiated sample.
In contrast, with a slight increase in fluence, reaching 0.65 dpa in maximum, we detect the onset of the $\beta$-to-$\gamma$ phase transition, featured by the formation of the $\beta$-$\gamma$ mixed layer [Fig.~\ref{fig:exp}(i)] sandwiched between the remaining $\beta$ film [Fig.~\ref{fig:exp}(b)]. %on the top of the sample and the $\beta$-substrate
For this sample, the $\beta$-$\gamma$ mixed layer consists of approximately 50/50 proportion, as shown in Fig.~\ref{fig:exp}(i).
Importantly, increasing disorder towards 0.85 dpa in maximum leads to the broadening of the mixed layer [Fig.~\ref{fig:exp}(c)] and an increase of the $\gamma$-phase fraction in this layer to $\leq 90\%$ [Fig.~\ref{fig:exp}(k)], revealing dominating \ce{Ga} planes with the order closely resembling that of $\gamma$-\ce{Ga}, with minor inclusions of \ce{Ga} planes with $\beta$-\ce{Ga} order, as shown as the region highlighted with pink color in Fig.~\ref{fig:exp}(k).
A further increase in dpa results in the formation of the homogeneous $\gamma$-layer on top of the $\beta$-phase substrate, as seen in Fig.~\ref{fig:exp}(d), (m), and (l), consistent with the literature~\cite{garcia2022formation, SFazarov2023universal}. 

Therefore, a disorder level in the range of 0.65–0.85 dpa is sufficient to, firstly, overcome the nucleation barrier and, secondly, stabilize the $\gamma$-phase at the specific temperature and dose rate used in this experiment (\textit{i.e.}, under conditions of a certain survival rate of the primary point damage interconnected with dpa).
The importance of post-cascade defect reactions becomes apparent when comparing the dpa values at the top $\beta$/$\gamma$ and bottom $\gamma$/$\beta$ interfaces for the samples in Fig.~\ref{fig:exp}(b) and (c).
A higher dpa value at the top $\beta$/$\gamma$ interface indicates that the disorder required for the phase transition depends on the proximity of the sample surface which acts as a sink for the radiation-induced defects.
Although further studies are needed to fully quantify the impact of secondary defect reactions, a clear correlation between the advancements of the top $\beta$/$\gamma$ and bottom $\gamma$/$\beta$ interfaces and the dpa values is observed in both samples in Fig.~\ref{fig:exp}(b) and (c).
Interestingly, in the sample implanted with the highest fluence used in this work, where a homogeneous $\gamma$-layer was formed, as seen in Fig.~\ref{fig:exp}(d), the dpa value at the bottom $\gamma$/$\beta$ interface is significantly higher than the dpa values at the same $\gamma$/$\beta$ interfaces in the samples irradiated with lower fluences, as shown in Fig.~\ref{fig:exp}(b) and (c).
This fact confirms that a non-linearity in the defect survival rate as a function of the cascade density~\cite{azarov2022interplay}.
Nevertheless, the overall trend governing the process is clear: the nucleation of the $\gamma$-phase starts upon reaching a disorder threshold, apparently in the maximum of the nuclear energy deposition region, corresponding to the depth where the dpa value is maximal, leading to its expansion towards the surface and into the bulk with increasing fluence.

\begin{figure}[ht!]
 \includegraphics[width=7cm]{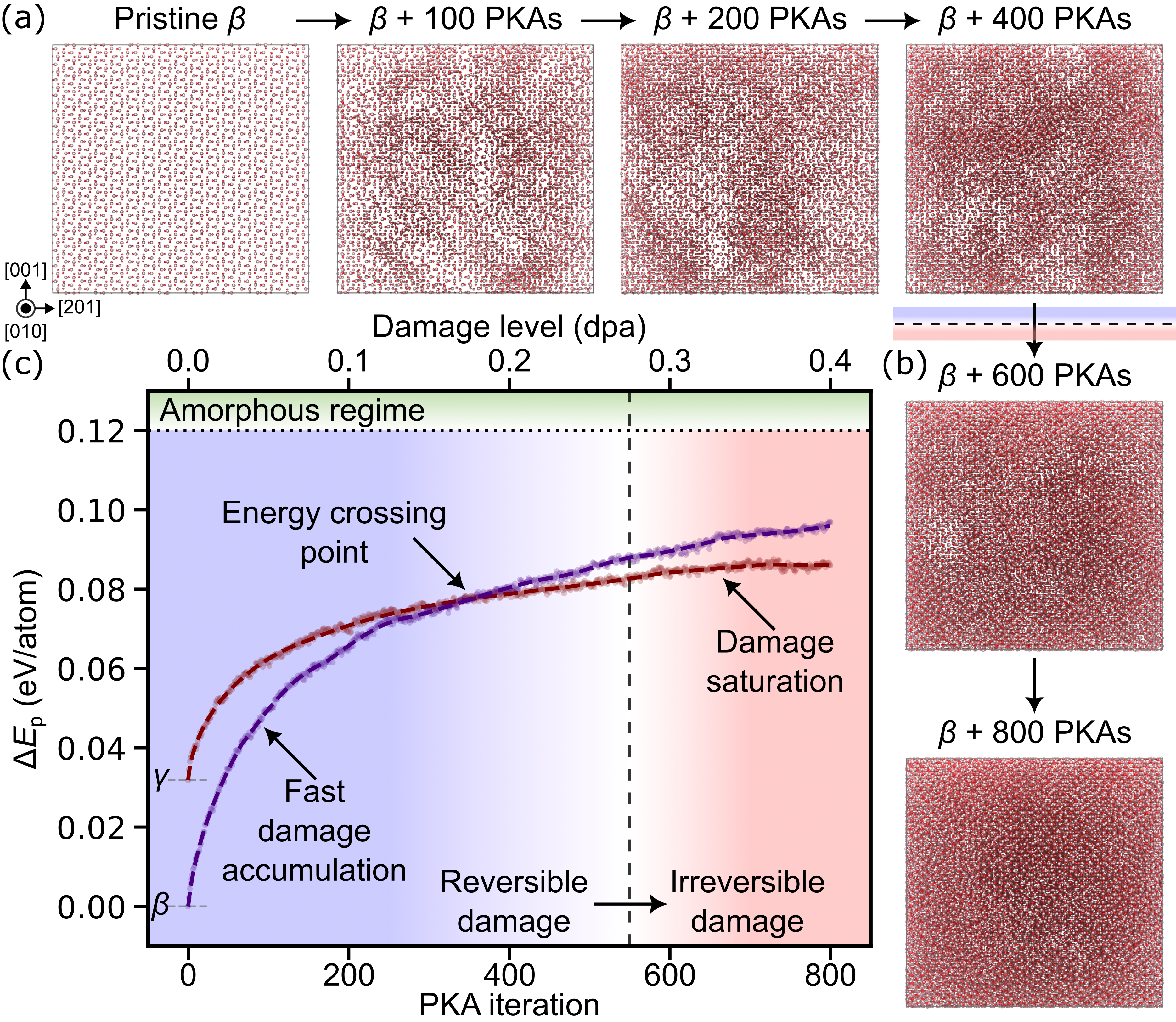}
 \caption{
 Structural evolution of $\beta$-\ce{Ga2O3} exposed to the overlapping cascade classified as (a) below the threshold reversible damage level, versus (b) above threshold of the irreversible transition damage level of 550 PKAs, $\sim 0.275$ dpa.
 The O and Ga atoms are in red and brown, respectively.
 (c) Potential energy, $\mathrm{\Delta} E_\mathrm{p}$, in $\beta$- and $\gamma$-phases as a function of PKAs normalized to the potential energy of the perfect $\beta$-phase as a zero-point.
 }
 \label{fig:irr}
\end{figure}

Importantly, these experimental observations are in excellent agreement with the results of our theoretical modeling. 
As shown in Fig.~\ref{fig:irr}, we employ ML-MD simulations to elucidate the underlying atomic-level mechanism of the swift $\beta$-to-$\gamma$ phase transition observed for the dpa range studied in Fig.~\ref{fig:exp}.
Commencing with the pristine $\beta$-\ce{Ga2O3} lattice in orthogonal cell (see SM Appendix II for computational details and construction of the orthogonal $\beta$ supercell), labelled as ``$\beta$'' in the rest of the Letter, the stochastic overlapping cascade process initially generates primary damage in the form of discrete point defect clusters [Fig.~\ref{fig:irr}(a)]. 
These defect clusters subsequently merge into a continuous disordered matrix [Fig.~\ref{fig:irr}(b)].
This structural evolution is clearly traceable through changes in the $\beta$-lattice pattern viewed from the monoclinic $\beta[010]$ orientation, which gradually disappears with the increasing number of primary knock-on atoms (PKAs). 
A distinct difference is observed between the $\beta$-cells with 400 and 600 PKAs [$\beta + 400$-PKAs and $\beta + 600$-PKAs cells, respectively, in Fig.~\ref{fig:irr}(a) and (b)] when the accumulated disorder completely fills the simulation cell.

\begin{figure*}[ht!]
 \includegraphics[width=13cm]{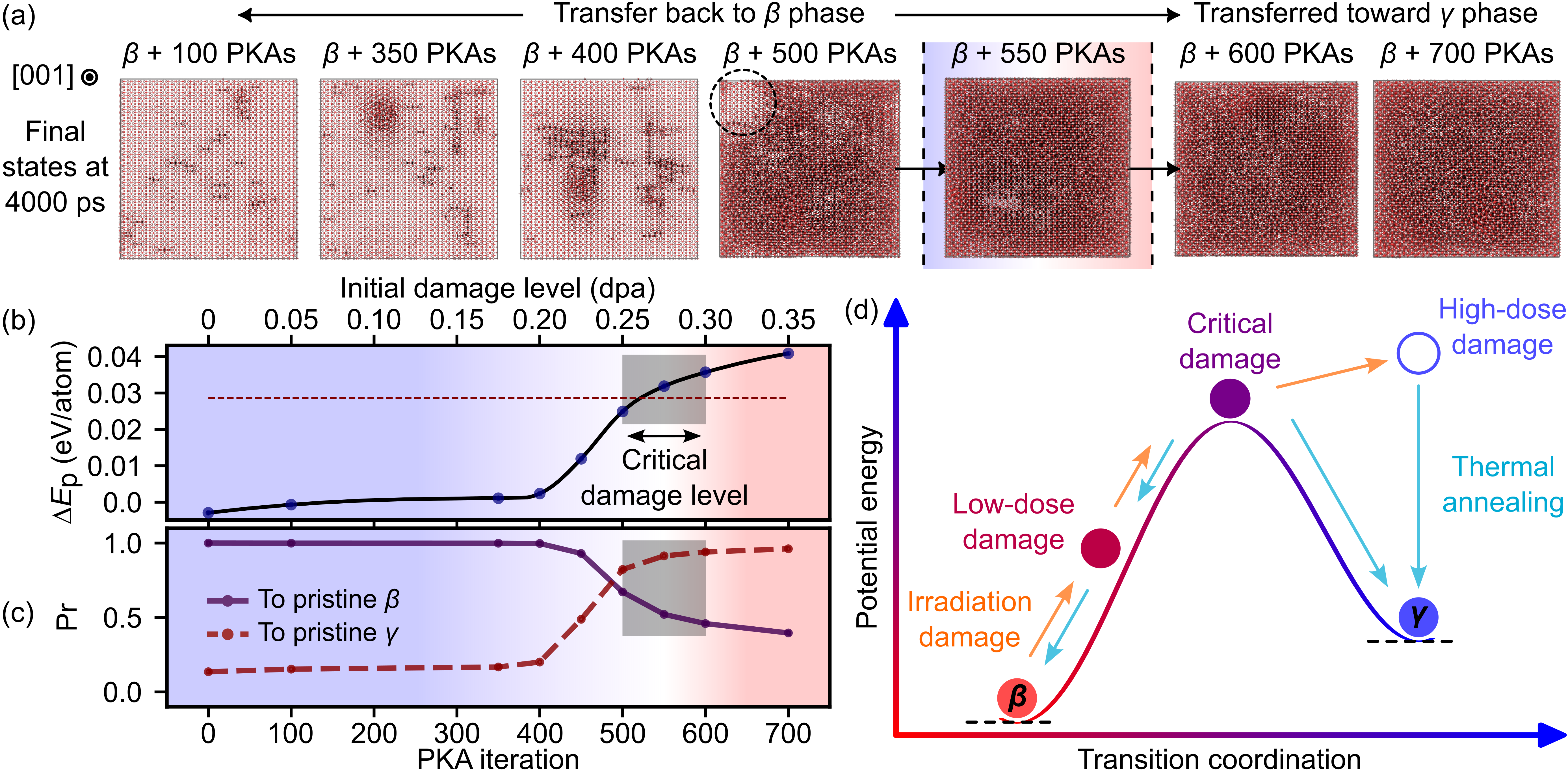}
 \caption{
 (a) Final states after 4000-ps annealing at 1500 K starting from different $\beta +$ PKAs states as indicated by the labels.
 The O and Ga atoms are in red and brown, respectively.
 (b) Final $\mathrm{\Delta} E_\mathrm{p}$ and (c) Pr values, upon 4000-ps anneals plotted versus PKAs pin-pointing the critical damage level for the irreversible transition. 
 See SM Appendix IV, Figs.~S7 and S8 for analyses on the PRDF and bond-angle distributions.
 (d) Schematics of the $\beta$-to-$\gamma$ phase transition mechanism in collision cascades.
 }
 \label{fig:homo_ann}
\end{figure*}

For quantitative analysis, we closely monitor the evolution of potential energy, $\mathrm{\Delta} E_\mathrm{p}$, as shown in Fig.~\ref{fig:irr}(c), comparing it to the $\mathrm{\Delta} E_\mathrm{p}$ evolution in similar simulations, but starting from a pristine $\gamma$-\ce{Ga2O3} cell (for detailed structural comparison and evolution, see SM Appendices III and IV, Figs.~S3 and S4).
Initially, the $\mathrm{\Delta} E_\mathrm{p}$ of $\gamma$-phase exhibits approximately 0.031 eV/atom higher value than that of the perfect $\beta$-phase.
The fast increasing trend in the $\mathrm{\Delta} E_\mathrm{p}$ curve of the $\beta$-phase during the early stages of the overlapping cascades (PKAs in the range of 0-200) signifies discrete primary damage accumulation. 
However, this trend slows down as newly generated defect clusters merge with the previously generated ones, leading to damage saturation. 
A similar trend is observed in the $\gamma$-phase curve, albeit with a significantly smaller gradient comparing to that for $\beta$-phase. 
This is consistent with the high radiation tolerance of the defective spinel structure of the $\gamma$-phase, which is fairly insensitive to the formation of new point defects in collision cascades~\cite{SFazarov2023universal}. 
The $\beta$- and $\gamma$-phases $\mathrm{\Delta} E_\mathrm{p}$ curves intersect after 350 PKAs, prominently correlating with the similarity analysis of the partial radial distribution functions (PRDFs) (SM Appendix IV, Fig.~S5).
After the crossing point, the $\mathrm{\Delta} E_\mathrm{p}$ of the $\beta$-phase is higher than that of $\gamma$-phase, which indicates that the consequent cascades disorder the $\beta$-phase more intensively than the $\gamma$-phase.
After 800 PKAs, the final $\mathrm{\Delta} E_\mathrm{p}$ values of the disordered cells increase by approximately 0.095 eV/atom and 0.050 eV/atom for the $\beta$- and $\gamma$-phases, respectively. 
Notably, at this high disorder level, both cells still exhibit significant energy differences (0.03-0.04 eV/atom) comparing to that of the amorphous \ce{Ga2O3} [marked as green zone in Fig.~\ref{fig:irr}(c)].     
Thus, for both phases, this disordering effect does not cause transition to amorphization, otherwise commonly observed in semiconductors under ion-beam irradiation~\cite{nord2002amorphization, pelaz2004ion}.
Importantly, the analysis of the PRDF and bond-angle distribution confirms that the $\beta$-\ce{O} sublattice retains its face-centered cubic lattice structure after 800 PKAs (SM Appendix IV, Figs.~S5 and S6).    

Furthermore, we investigated the structural evolution in the $\beta$-phase cells due to the relaxation of the damage accumulated at different dpa levels. 
We employ annealing simulations at 1500 K and 0 bar, as summarized in Fig.~\ref{fig:homo_ann}. 
Fig.~\ref{fig:homo_ann}(a) illustrates seven distinct configurations observed after the 4000-ps annealing, as viewed from $\beta[001]$ direction.
In Fig.~\ref{fig:homo_ann}(b), we display the corresponding final $\mathrm{\Delta} E_\mathrm{p}$ with respect to a zero level corresponding to the pristine $\beta$-phase, and the brown line indicating the pristine $\gamma$-phase under the same annealing conditions (1500 K and 0 bar).
The PRDFs of the second \ce{Ga}-\ce{Ga} shell is used for the similarity analysis. 
In Fig.~\ref{fig:homo_ann}(c), the Pearson correlation coefficient, $\mathrm{Pr}$, is calculated with respect to the pristine $\beta$-\ce{Ga} (purple line) and $\gamma$-\ce{Ga} (brown lines) PRDFs as a function of the PKA iteration.
See SM Appendix IV, Figs.~S7 and S8 for analyses on the PRDF and bond-angle distributions of the final states.
Combining the lattice configurations, the final $\mathrm{\Delta} E_\mathrm{p}$, and $\mathrm{Pr}$ data in Fig.~\ref{fig:homo_ann}(a-c), the final states can be grouped into three categories: 
(\textit{i}) $\beta$-phase with residual defect clusters observed after 100/350/400 PKAs; 
(\textit{ii}) critical transition region with combined $\beta$/$\gamma$/disordered phases, observed after 500/550/600 PKAs; 
and (\textit{iii}) combined $\gamma$/disordered phases which can be observed after 700 PKAs.
For example, annealing of the $\beta + 500$-PKAs state results in recovery of a significant fraction of the $\beta$-phase, as indicated by the dashed circle in Fig.~\ref{fig:homo_ann}(a).

\begin{figure*}[ht!]
 \includegraphics[width=13cm]{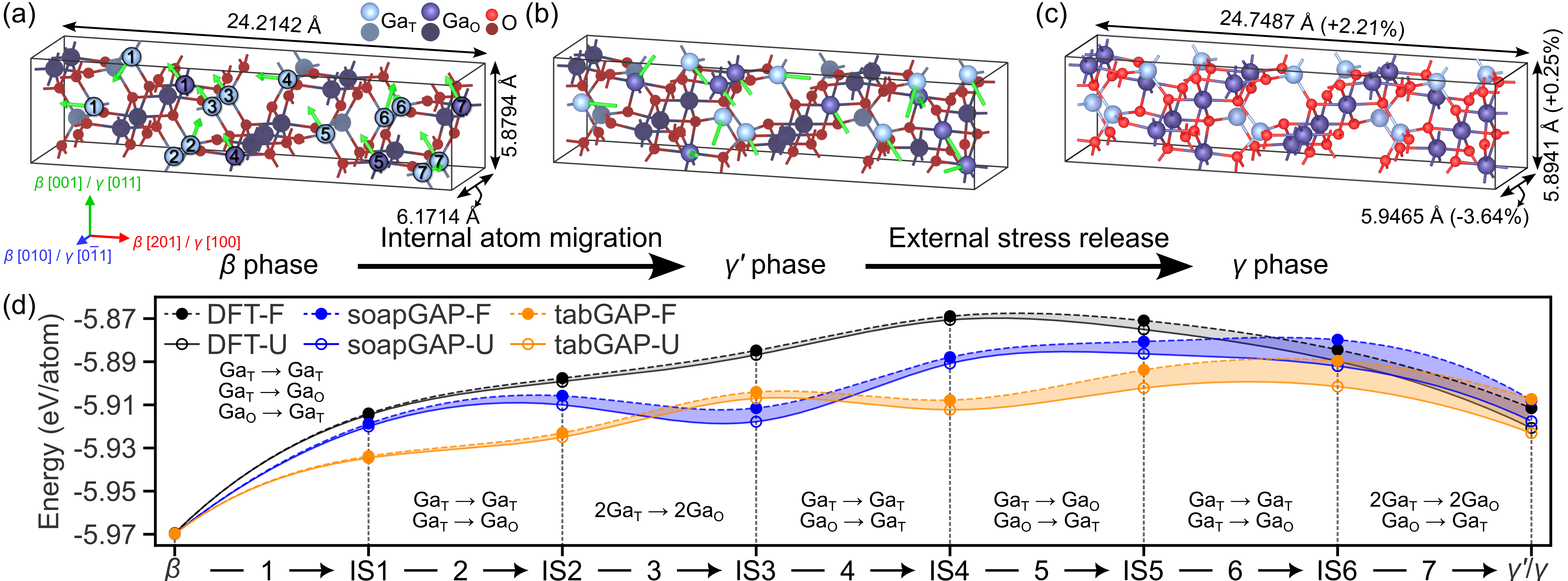}
 \caption{
 % Complete atomic migration pathways of $\beta$-to-$\gamma$ \ce{Ga2O3} phase transition with full periodic boundary.
 (a) An initial 80-atom orthogonal $\beta$-phase supercell. 
 The tetrahedral \ce{Ga} (\ce{Ga}$_\mathrm{T}$), octahedral \ce{Ga} (\ce{Ga}$_\mathrm{O}$), and \ce{O} atoms are colored in light blue, dark blue, and red, respectively. 
 The 16 migrating \ce{Ga} atoms (12 \ce{Ga}$_\mathrm{T}$ and 4 \ce{Ga}$_\mathrm{O}$) are highlighted and labeled with the indices of migration sequence from 1 to 7, corresponding to the indices shown in (d). 
 The displacement vectors are illustrated with green arrows. 
 (b) The resultant $\gamma'$ phase after internal \ce{Ga} atom migration, conserving the shape of the $\beta$ cell.
 (c) The final $\gamma$ phase after external stress release.  
 (d) The potential energy landscape of the pathways, probed using DFT, ML-soapGAP, and ML-tabGAP, from $\beta$ to $\gamma'/\gamma$ phases with six intermediate states (IS1-IS6).
 % The potential energies of the intermediate states are probed using DFT, ML-soapGAP, and ML-tabGAP by optimizing the systems with the fixed (F) $\beta$-cell side lengths and the unfixed (U) side lengths to release external stress.  
 }
 \label{fig:atomic_path}
\end{figure*}

As summarized in Fig.~\ref{fig:homo_ann}(b) and (c), the final $\mathrm{\Delta} E_\mathrm{p}$ and $\mathrm{Pr}$ values reached by the systems after the 4000-ps annealing exhibit a sharp change starting in the range of $450-500$ PKAs.
Consequently, in accordance with the analysis in Fig.\ref{fig:irr}, the initial damage level of $0.25-0.3$ dpa is interpreted as the critical disorder threshold.
Thus, the simulated dpa-level required for the irreversible $\beta$-to-$\gamma$ transition aligns remarkably well with the experimental/SRIM data in Fig.~\ref{fig:exp}.     
At this end, the mechanism of the irradiation-induced $\beta$-to-$\gamma$ transition is illustrated by a schematics in Fig.~\ref{fig:homo_ann}(d).
The irradiation-induced disorder primarily introduces \ce{Ga} defect clusters which reorder during post-cascade periods. 
At low-dpa levels, these \ce{Ga} defect clusters are sparsely embedded into the $\beta$-phase and can swiftly recover due to rapid migration of \ce{Ga} atoms. 
Conversely, the critical damage level is reached when the accumulated \ce{Ga} defect clusters completely replace the original $\beta$-phase. 
In this state, the system with high potential energy becomes rather unstable and can only exist transiently.
Instead of reverting to the low-symmetry ($C2/m$) $\beta$-lattice, the \ce{Ga} sublattice kinetically favors transitioning to the metastable cubic ($Fd\overline{3}m$) $\gamma$-lattice, requiring significantly less migration of \ce{Ga} atoms to reach the lattice sites.

Figs.~\ref{fig:exp}-\ref{fig:homo_ann} elucidate the $\beta$-to-$\gamma$ phase transition in large-scale stochastic dynamical systems.
However, it is instructive to isolate two otherwise coupled processes -- internal atom migration and external stress release -- to present the kinetically minimized atomic migration pathways in the \ce{Ga} sublattice, as shown in Fig.~\ref{fig:atomic_path}. 
An 80-atom orthogonal supercell is used as the smallest cell to illustrate the overall symmetry transition of the initial $C2/m$ $\beta$-\ce{Ga} [Fig.~\ref{fig:atomic_path}(a)]. 
Remarkably, short-range migrations involving precisely half of the total \ce{Ga} atoms (16 out of 32) complete the internal atomic transition.
These \ce{Ga} atoms are highlighted in Fig.~\ref{fig:atomic_path}(a) and indexed based on the migration sequences in Fig.~\ref{fig:atomic_path}(d), resulting in a minimal effective dpa of 0.20 (16 displaced atoms out of 80 atoms in total), aligning excellently with the critical dpa level (0.25-0.30) in Fig.~\ref{fig:homo_ann}(d).
In total, 12 tetrahedral \ce{Ga} (\ce{Ga}$_\mathrm{T}$) and 4 octahedral \ce{Ga} (\ce{Ga}$_\mathrm{O}$) atom displacements yield 8 new \ce{Ga}$_\mathrm{T}$ and 8 \ce{Ga}$_\mathrm{O}$. 
Following the displacement, the \ce{Ga}$_\mathrm{T}$/\ce{Ga}$_\mathrm{O}$ ratio changes from $1:1$ to $3:5$, with 4 unoccupied \ce{Ga}$_\mathrm{O}$ sites completing a defect-free spinel \ce{Ga3O4} lattice. 
The intermediate state that we label as $\gamma'$-phase displays the typical hexagonal projected pattern of the $\gamma$-\ce{Ga} sublattice when viewed from the $\gamma [011]$ and $\gamma [0\overline{1}1]$ directions (SM Appendix V, Fig. S9, and Video S1).    
Nevertheless, the $\gamma'$-phase retains the $\beta$-\ce{Ga2O3} cell side-length, leading to a non-uniform external stress (and strain) accumulating in the system. 
Thus, the phase transition is further finalized via cell relaxation, associated with marginal internal atom rearrangement at local sites, as shown in Fig.~\ref{fig:atomic_path}(c).
This scenario is in good agreement with the experimental observation of the strain release accompanying with the $\beta$-to-$\gamma$ phase transition measured by the SAED~\cite{azarov2022disorder, garcia2022formation, SFazarov2023universal}.

Even though the decoupling above is instructive, in real material system, internal atom migration and external stress release naturally occur in combination during phase transitions.
Therefore, in Fig.~\ref{fig:atomic_path}(d), the energy evolution of the fixed (F) and unfixed (U) cells is calculated using density functional theory (DFT) and the two ML models to map the intermediate potential energy landscape.
The overall potential energy trends predicted by all three models are comparable, with the highest potential energy reaching approximately 0.12 eV/atom above the potential energy of the $\beta$-phase.   
The overall transition pathways can be divided into 7 migration sequences with six intermediate states [IS1-IS6 in Fig.~\ref{fig:atomic_path}(d)]. 
Each migration sequence involves two or three \ce{Ga} atom displacements, and consistently with the decoupling, as discussed above, significant stress release (indicated by the energy difference between the fixed and unfixed cells) is only observed after a majority of the \ce{Ga} have been displaced [from IS4 to $\gamma'/\gamma$ in Fig.~\ref{fig:atomic_path}(d)].
In this context, the fact that the disorder threshold required for $\beta$-to-$\gamma$ phase transition is empirically found consistently close to 0.2 dpa -- see Figs.~\ref{fig:irr} and \ref{fig:homo_ann} -- is not a coincidence, but is a natural consequence of the transition pathways allowing the minimal kinetics with the lowest disorder threshold (see Fig.~\ref{fig:atomic_path}).
In its turn, while comparing with experiment, the recombination of defects and repetitive atom displacement may be the factors requiring disorders larger than 0.2 dpa for making the $\beta$-to-$\gamma$ phase transition sustainable. 
% Nevertheless, the overall agreement among our experiments, MD simulations and theoretical pathways is excellent.   

In this Letter, we study a rare phenomenon of the crystallization instead of amorphization in collision cascades in $\beta$-\ce{Ga2O3}.
Specifically, the amorphization is suppressed by the polymorphic $\beta$-to-$\gamma$ phase transition.%, and the atomic-level mechanism is revealed. 
We accurately determine the critical disorder level that triggers the transition from disorder state to $\gamma$-phase. 
We show that below this threshold, the high-energy $\gamma$-\ce{Ga2O3} forms only transiently. 
However, upon reaching the threshold, instead of reverting to the low-symmetry $\beta$-phase, the \ce{Ga}-sublattice kinetically favors transitioning to the defective spinel $\gamma$-phase, requiring significantly less migration for \ce{Ga} atoms to reach the $\gamma$-lattice sites during the post-cascade processes. 
Moreover, we described full atomic-level migration pathways of $\beta$-to-$\gamma$ phase transition involving only short-distance $\beta$-\ce{Ga} displacements together with external stress release. 
% Leveraging on the state-of-the-art atomic-resolution STEM and ML-MD techniques, 
% In a broader perspective, understanding this remarkable phase transition mechanism paves the way for the atom-level phase engineering with ion beams, potentially enabling innovative device design strategies.

% \newpage


\end{document}